\documentclass[12pt,a4paper, amsmath,amssymb]{article}
\usepackage[margin=0.8in]{geometry}
\usepackage{longtable}
\usepackage{amssymb}
\usepackage{amsmath}
\usepackage{amsfonts}
\usepackage{bm}
\usepackage[utf8]{inputenc}
\usepackage{graphicx}
\usepackage{float}
\usepackage{xcolor}
\definecolor{bluemunsell}{rgb}{0.0, 0.5, 0.69}
\usepackage[colorlinks,linkcolor = bluemunsell,
urlcolor  = bluemunsell,
citecolor = bluemunsell,
anchorcolor = bluemunsell]{hyperref}
\usepackage{MnSymbol,bbding,pifont}
\usepackage{xcolor}
\usepackage{makecell}
\usepackage{ctable} 
\usepackage{adjustbox}

\usepackage{cite}
\usepackage{setspace}
\usepackage{caption}
\newcommand{\beq}{\begin{eqnarray}}
	\newcommand{\eeq}{\end{eqnarray}}

\setcellgapes{0.5pt}

\def\beqa{\begin{eqnarray}}
	\def\eeqa{\end{eqnarray}}

\evensidemargin \oddsidemargin
\renewcommand{\arraystretch}{1.5}

\let\OLDthebibliography\thebibliography
\renewcommand\thebibliography[1]{
	\OLDthebibliography{#1}
	\setlength{\parskip}{0pt}
	\setlength{\itemsep}{0pt plus 0.3ex}}
\usepackage{makecell}

\newcommand{\GeV}{{\rm\ GeV}}
\newcommand{\TeV}{{\rm\ TeV}}

\def\bsp#1\esp{\begin{split}#1\end{split}}
\def\bpm{\begin{pmatrix}}
	\def\epm{\end{pmatrix}}

\usepackage[normalem]{ulem}

\graphicspath{{Figures/}} %

\newcommand{\orcid}[1]{\hspace{1mm}\href{https://orcid.org/#1}{\includegraphics[height=0.3cm,keepaspectratio]{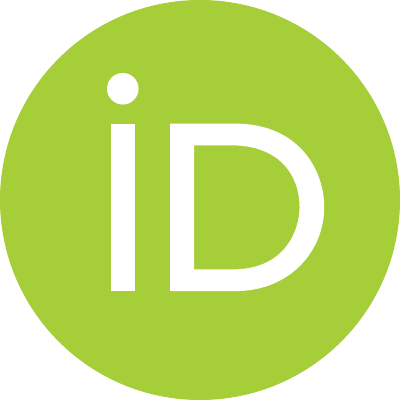}}}
\begin{document}
\begin{center}
{\Large \textbf{Unified Interpretation of  95 GeV Excesses in the Two Higgs Doublet type II Seesaw Model\\ 
}}
\thispagestyle{empty} 
\vspace{1cm}
{\bf
	Brahim Ait-Ouazghour\orcid{0009-0006-1419-969X}$^1$\footnote{\href{b.ouazghour@gmail.com}{b.ouazghour@gmail.com}},
	Mohamed Chabab\orcid{0000-0002-2772-4290}$^{1,2}$\footnote{\href{mchabab@uca.ac.ma (Corresponding author)}{mchabab@uca.ac.ma (Corresponding author)}},
	Khalid Goure\orcid{0009-0007-5292-5012}$^1$\footnote{\href{khalidgoure01@gmail.com }{khalidgoure01@gmail.com}} \\
}
\vspace{0.4cm}
{\small
	$^1$LPHEA, Faculty of Science Semlalia, Cadi Ayyad University, P.O.B. 2390 \\ Marrakech, Morocco. \\
	\small  $^2$ National School of Applied Science,  P.O.B 63 Saffi 46000, Morocco
}

\end{center}
\begin{abstract}
In the search for a light Higgs boson, the ATLAS and CMS experiments have observed excesses in both the diphoton ($\gamma\gamma$) and di-tau-pair ($\tau^+\tau^-$) decay channels at about $95$ GeV. The LEP collaboration has also previously reported an excess in the $b\bar{b}$ channel  at a comparable Higgs mass. In this paper, we explore whether these excesses can be accommodated within the framework of the Two Higgs Doublet type II Seesaw Model (2HDMcT). By implementing various theoretical constraints and experimental limits on the parameter space, we first demonstrate that a light CP-even Higgs boson, $h_1$, with a mass around 95 GeV can simultaneously account for the excesses observed in the $\gamma\gamma$ and $b\bar{b}$ channels, provided a Type I Yukawa texture is employed. More interestingly, our analysis shows that the three excesses in  $\gamma\gamma$, $b\bar{b}$ and $\tau^+\tau^-$ channels can well be accommodated simultaneously, reaching a $0.64 \sigma$ C.L. if the CP-odd Higgs boson $A_1$ is nearly mass degenerate and superposed to the light CP-even Higgs $h_1$.
\end{abstract}


	\section{Introduction}

\small
The discovery of the $125$ GeV Higgs boson by the ATLAS and CMS collaborations \cite{ATLAS:2012yve,CMS:2012qbp,ATLAS:2016neq} marks an important milestone in the quest to unravel the nature of electroweak symmetry breaking (EWSB). This discovery established the Standard Model (SM) as a robust and  successful framework for describing the fundamental particles and their interactions. Though the SM predictions have been rigorously tested \cite{CMS:2022dwd,ATLAS:2022vkf}, it still faces challenges since it failed to account for several physical phenomena, such as: the origin of dark matter \cite{Zwicky:1933gu,Rubin:1970zza}, the hierarchy problem \cite{Veltman:1980mj}, and the neutrino mass generation \cite{RevModPhys.88.030501,RevModPhys.88.030502}. To tackle these deficiencies, numerous theoretical frameworks, beyond the Standard Model (BSM) have been suggested in the literature.  BSM scenarios with an extended Higgs sector are among the most popular ones. The fermion sector of the SM being non minimal, it is legitimate to suggest scenarios with an extended scalar sector, hence a rich Higgs spectrum, aiming to  accommodate new physics. Further investigation of EWSB, precise determination of the properties of the observed $125$ GeV Higgs and the searches for new scalar particles are actively conducted at the LHC.  Among important key focus of the recent experimental physics programs are the observation of  additional Higgs bosons with light mass below $125 \GeV$ LEP~\cite{OPAL:2002ifx,LEPWorkingGroupforHiggsbosonsearches:2003ing,ALEPH:2006tnd}, Tevatron~\cite{CDF:2012wzv} and  LHC~\cite{CMS:2015ocq,CMS:2018cyk,CMS-PAS-HIG-21-001,CMS-PAS-HIG-17-013, CMS:2018rmh,CMS-PAS-HIG-20-002,ATLAS-CONF-2018-025,CMS:2024yhz,ATLAS:2024bjr}. These searches have been performed on different Higgs decay channels:  based on the $8 \TeV$ data and Run~2  first year data at $13 \TeV$, corresponding to an integrated luminosity of $19.7\,\mathrm{fb}^{-1}$ and $35.9\,\mathrm{fb}^{-1}$, respectively, CMS results  first revealed a local excess of $2.8\, \sigma$ at $95.3 \GeV$~\cite{CMS:2018cyk}. Later, the CMS collaboration  using the full Run 2 data set, reported  a local excess of $2.9\ \sigma$ at $95.4$ GeV \cite{CMS-PAS-HIG-20-002,CMS:2024yhz}. Similarly the ATLAS experiment searching for diphoton resonances using $140\,\mathrm{fb}^{-1}$ of pp collisions, found a local excess of $1.7\,\sigma$ around $95$ GeV  \cite{ATLAS-CONF-2023-035,ATLAS:2024bjr}. Besides the di-photon excess, another excess in the $\tau^-\tau^+$ channel has been observed by CMS \cite{CMS:2022goy}. Notably, many years ago, LEP also reported a comparable local $2.3,\sigma$ excess in the $e^+e^- \to Z(\phi \to b\bar{b})$ channel \cite{LEPWorkingGroupforHiggsbosonsearches:2003ing}. So, these observations deserve  further phenomenological investigations within BSM models, in particular, to see  wether the excesses can be interpreted via scalar particles at about $95$ GeV in their extended Higgs sectors.  Many attempts have been performed in different theoretical frameworks  ~\cite{Cao:2016uwt,Cacciapaglia:2016tlr,Crivellin:2017upt,Biekotter:2019kde,Cline:2019okt,Cao:2019ofo,Abdelalim:2020xfk,Biekotter:2021ovi,Heinemeyer:2021msz,Biekotter:2021qbc,Biekotter:2022abc,Iguro:2022dok,Li:2022etb,Biekotter:2022jyr,Biekotter:2023jld,Azevedo:2023zkg,Biekotter:2023oen,Ahriche:2023hho,Chen:2023bqr,Cao:2024axg,Arhrib:2024wjj,Wang:2024bkg,Li:2023kbf,Dev:2023kzu,Borah:2023hqw,Cao:2023gkc,Ellwanger:2023zjc,Aguilar-Saavedra:2023tql,Ashanujjaman:2023etj,Dutta:2023cig,Ellwanger:2024txc,Diaz:2024yfu,Ellwanger:2024vvs,Ayazi:2024fmn,Lian:2024smg,Yang:2024fol,Ge:2024rdr,YaserAyazi:2024hpj,Janot:2024ryq,Mosala:2024mcy,Gao:2024qag,Kalinowski:2024uxe,Benbrik:2024ptw,Khanna:2024bah}.

In this  work, our analysis is done within the framework of the Two Higgs Doublet Type II seesaw Model (2HDMcT). Besides the two Higgs doublets with hypercharge $Y=1$, The Two Higgs Doublet Model (2HDM) is augmented with an $SU(2)_L$ complex scalar triplet $\Delta$ with $Y_\Delta=2$. Due to the similarity between mass generation from the seesaw mechanism and the Brout-Englert-Higgs mechanism, the 2HDMcT model is highly attractive: it exhibits numerous phenomenological features, particularly distinct from those arising in the scalar sector of the 2HDM. Apart its broader spectrum in comparison to $2HDM$, it is arguably among the simplest and well motivated frameworks that can explain either  all neutrino oscillation data in a gauge-invariant way \cite{FileviezPerez:2008jbu,Arhrib:2011uy,King:2015aea,Cai:2017mow}, or the dark matter origin \cite{Chen:2014lla}. In addition to that, the doubly charged Higgs $H^{++}$ is considered a  smoking gun of 2HDMcT, $H^{++}$ being intensively searched for at ATLAS and CMS  through promising  $H^{++}H^{--}$ and $H_i^{+}H_i^{-}$ $(i=1,2)$ channels decaying to the same sign di-lepton \cite{Ouazghour:2018mld}. Moreover beyond the Higgs phenomenology, it has been demonstrated that interactions between doublet and triplet fields in this model can induce a strong first order electroweak phase transition, hence providing conditions for the generation of the baryon asymmetry through electroweak baryongenesis \cite{Ramsey-Musolf:2019lsf}.  Furthermore,  this model provides other appealing features including possibility of enhanced Higgs couplings, modified Higgs decay channels, and the presence of additional scalar particles that can be probed at current or  future colliders. As an example, besides  contributions from SM charged particles ($W^{\pm}$, and fermions), the loop-mediated decay $h_1\to \gamma\gamma$ receives extra contributions from the  singly and doubly charged Higgs bosons, predicted by 2HDMcT spectrum \cite{Ouazghour:2024fgo}. This could  increase the branching ratio ${\cal BR}_{\rm 2HDMcT}(h_1\to \gamma\gamma)$ (For more details see Appendix \ref{appendice:A}). Thus the aim of this paper is to investigate whether the 2HDMcT, with its broader spectrum compared to simpler models, can account for the reported excesses \cite{CMS:2024yhz,ATLAS:2024bjr,LEPWorkingGroupforHiggsbosonsearches:2003ing}.
\noindent

To perform a rigorous study, we have implemented a full set of theoretical constraints originated from perturbative unitarity, and electroweak vacuum stability. To accommodate the $125$~GeV Higgs boson $h_{125}$ we implement the \texttt{HiggsTools} package~\cite{Bahl:2022igd} subpackage \texttt{HiggsSignals}~\cite{Bechtle:2013xfa,Bechtle:2014ewa,Bechtle:2020uwn,Bahl:2022igd}. Also to  guarantee that the allowed parameter space aligns with exclusion limits resulting from searches for additional Higgs bosons at the LHC and LEP, the subpackage \texttt{HiggsBounds}~\cite{Bechtle:2008jh,Bechtle:2011sb,Bechtle:2013wla,Bechtle:2020pkv,Bahl:2022igd} is employed. In addition, the 2HDMcT parameter space is further constrained by \texttt{electroweak precision observables} and the \texttt{$\bar{B}\to X_{s}\gamma$ constraint} at 95$\%$ C.L.

The paper is organized as follows. In Sec.\ref{section2}, we briefly introduce the 2HDMcT, specifically focusing on the CP-even scalar spectrum. In Sec.\ref{section3}, we first provide a brief discussion of the relevant theoretical and experimental constraints delimiting  our  model parameter space, then  we present our analysis and results. Sect.\ref{conclusion} is devoted to our conclusion.

\section{The model: 2HDMcT }
\label{section2}

2HDMcT includes two Higgs doublets with a hypercharge $Y=+1$ and one complex scalar triplet
with $Y=+2$. After  the electroweak
symmetry is spontaneously broken, these fields can be parameterized as :
\begin{equation}
	\Phi_1 = \left(
	\begin{array}{c}
		\phi_1^+ \\
		\phi_1^0 \\
	\end{array}
	\right)
	\quad {\rm ,}\quad 
	\Phi_2 = \left(
	\begin{array}{c}
		\phi_2^+ \\
		\phi_2^0 \\
	\end{array}
	\right)
		\quad {\rm ,}\quad 
		\begin{matrix}
			\Delta &=&\left(
			\begin{array}{cc}
				\delta^+/\sqrt{2} & \delta^{++} \\
				(v_t+\delta^0+i\eta_0)/\sqrt{2} & -\delta^+/\sqrt{2}\\
			\end{array}
			\right)
				\end{matrix}
\end{equation}

with $\phi_1^0 = (v_1+\psi_1+ i \eta_1)/\sqrt{2}$, $\phi_2^0 = (v_2+\psi_2+ i \eta_2)/\sqrt{2}$ and $\sqrt{v_1^2+v_2^2+2v_t^2}=246$ GeV, where $v_1$,$v_2$ and $v_t$ are the vacuum expectation values of the two Higgs doublet and triplet fields respectively.

\noindent
The most general $SU(2)_L\times U(1)_Y$ invariant Lagrangian of 2HDMcT is given by:
{\begin{equation}
		\begin{matrix}
			\mathcal{L}=\sum_{i=1}^2(D_\mu{\Phi_i})^\dagger(D^\mu{\Phi_i})+Tr(D_\mu{\Delta})^\dagger(D^\mu{\Delta})\vspace*{0.12cm}\\
			\hspace{-3cm}-V(\Phi_i, \Delta)+\mathcal{L}_{\rm Yukawa}
			\label{eq:thdmt-lag}
		\end{matrix}
\end{equation}}
where the covariant derivatives are defined as,
\begin{equation}
	D_\mu{\Phi_i}=\partial_\mu{\Phi_i}+igT^a{W}^a_\mu{\Phi_i}+i\frac{g'}{2}B_\mu{\Phi_i} \label{eq:covd1}
\end{equation}

\begin{equation}
	D_\mu{\Delta}=\partial_\mu{\Delta}+ig[T^a{W}^a_\mu,\Delta]+ig' \frac{Y_\Delta}{2} B_\mu{\Delta} \label{eq:covd2}
\end{equation}
(${W}^a_\mu$, $g$), and ($B_\mu$, $g'$) denoting respectively the $SU(2)_L$ and $U(1)_Y$ gauge fields and couplings and $T^a \equiv \sigma^a/2$, with $\sigma^a$ ($a=1, 2, 3$)  the Pauli matrices. In terms of the two $SU(2)_L$ Higgs doublets $\Phi_i$ and the triplet field $\Delta$, the 2HDMcT scalar potential is given by \cite{Ouazghour:2018mld, Chen:2014xva}:
	\[
	\begin{matrix}
		V(\Phi_1,\Phi_2,\Delta) &=& m_{11}^2 \Phi_1^\dagger\Phi_1+m_{22}^2\Phi_2^\dagger\Phi_2-[m_{12}^2\Phi_1^\dagger\Phi_2+{\rm h.c.}]+\frac{\lambda_1}{2}(\Phi_1^\dagger\Phi_1)^2
		+\frac{\lambda_2}{2}(\Phi_2^\dagger\Phi_2)^2
		\nonumber\\
		&+& \lambda_4(\Phi_1^\dagger\Phi_2)(\Phi_2^\dagger\Phi_1)+ \left\{\frac{\lambda_5}{2}(\Phi_1^\dagger\Phi_2)^2
		+\big[\beta_1(\Phi_1^\dagger\Phi_1)
		+\beta_2(\Phi_2^\dagger\Phi_2)\big]
		\Phi_1^\dagger\Phi_2+{\rm h.c.}\right\} \nonumber\\
		&+&\lambda_3(\Phi_1^\dagger\Phi_1)(\Phi_2^\dagger\Phi_2)+\lambda_6\,\Phi_1^\dagger \Phi_1 Tr\Delta^{\dagger}{\Delta} +\lambda_7\,\Phi_2^\dagger \Phi_2 Tr\Delta^{\dagger}{\Delta}\nonumber\\
		&+&\left\{\mu_1 \Phi_1^T{i}\sigma^2\Delta^{\dagger}\Phi_1 + \mu_2\Phi_2^T{i}\sigma^2\Delta^{\dagger}\Phi_2 + \mu_3 \Phi_1^T{i}\sigma^2\Delta^{\dagger}\Phi_2  + {\rm h.c.}\right\}+\lambda_8\,\Phi_1^\dagger{\Delta}\Delta^{\dagger} \Phi_1\nonumber\\
		&+&\lambda_9\,\Phi_2^\dagger{\Delta}\Delta^{\dagger} \Phi_2+m^2_{\Delta}\, Tr(\Delta^{\dagger}{\Delta}) +\bar{\lambda}_8(Tr\Delta^{\dagger}{\Delta})^2\hspace*{0cm}+\hspace*{0cm}\bar{\lambda}_9Tr(\Delta^{\dagger}{\Delta})^2
		\label{scalar_pot}
	\end{matrix}
	\]
In this work , we avoid tree-level Higgs mediated $FCNC_s$ by considering $Z_2$ symmetry ($\Phi_1\rightarrow+\Phi_1$ and $\Phi_2\rightarrow-\Phi_2$) with $\beta_1=\beta_2=0$ . The $Z_2$ symmetry is softly broken by the bi-linear terms proportional to $m_{12}^2$, $\mu_1$, $\mu_2$ and $\mu_3$ parameters. In addition, we assume that $m_{11}^2$, $m_{22}^2$, $m_{\Delta}^2$, $m_{12}^2$, $\lambda_{i}$ $(i=1,..,9)$, $\bar{\lambda}_{j}$ $(j=8,9)$, $\mu_{k}$ $(k=1,..,3)$, $\beta_{1,2}$ are real potential parameters.   
In the CP-even scalar sector, the mixing of the states ($\rho_{1},\rho_{2},\rho_{0}$) leads to a total of three CP-even physical Higgs bosons($h_1,h_2,h_3$). The neutral scalar mass matrix reads as,
\noindent
\begin{eqnarray}
	{\mathcal{M}}_{{\mathcal{CP}}_{even}}^2=\left(
	\begin{array}{ccc}
		m_{\rho_{1}\rho_{1}}^2  &  m_{\rho_{2}\rho_{1}}^2  &  m_{\rho_0\rho_{1}}^2   \\
		m_{\rho_{1}\rho_{2}}^2  &  m_{\rho_{2}\rho_{2}}^2  &  m_{\rho_0\rho_{2}}^2   \\
		m_{\rho_{1}\rho_0}^2  &  m_{\rho_{2}\rho_0}^2  &  m_{\rho_0\rho_0}^2      
	\end{array}
	\right)
	\label{matrix-cp-even}   
\end{eqnarray}
\noindent
where the diagonal terms are given by,
\noindent
\begin{eqnarray}
	&&m_{\rho_{1}\rho_{1}}^2 = \lambda_1 v_1^2+\frac{v_2 \left(\sqrt{2} m_{3}^2 + \mu_3 v_t\right)}{\sqrt{2} v_1} \nonumber\\
	&&m_{\rho_{2}\rho_{2}}^2 = \lambda_2 v_2^2+\frac{v_1 \left(\sqrt{2} m_{3}^2+ \mu_3 v_t\right)}{\sqrt{2} v_2} \nonumber\\
	&&m_{\rho_0\rho_0}^2 = \frac{4 \left(\bar{\lambda}_{8}+\bar{\lambda}_{9}\right) v_t^3+\sqrt{2} \left(\mu_1 v_1^2+\mu_3 v_2 v_1+\mu_2 v_2^2\right)}{2 v_t}
	\label{diago-cp-even}
\end{eqnarray}
\noindent
while the off-diagonal terms are,
%
\begin{eqnarray}
	&&m_{\rho_{2}\rho_{1}}^2 = m_{\rho_{1}\rho_{2}}^2 = \frac{1}{\sqrt{2}} \left(\sqrt{2} v_1 v_2 \lambda_{345} - \sqrt{2} m_{3}^2-\mu_3 v_t\right) \nonumber\\
	&&m_{\rho_0\rho_{1}}^2 = m_{\rho_{1}\rho_0}^2 = \frac{1}{\sqrt{2}} \left(\sqrt{2} v_1 v_t (\lambda_6+\lambda_8) - (2 \mu_1 v_1 + \mu_3 v_2)\right) \nonumber\\
	&&m_{\rho_0\rho_{2}}^2 = m_{\rho_{2}\rho_0}^2 = \frac{1}{\sqrt{2}} \left(\sqrt{2} v_2 v_t (\lambda_7+\lambda_9) - (2 \mu_2 v_2 + \mu_3 v_1)\right) 
	\label{off-diago-cp-even} 
\end{eqnarray}
${\mathcal{M}}_{{\mathcal{CP}}_{even}}^2$ mass matrix can be diagonalized by an orthogonal matrix ${\mathcal{E}}$ parametrized as,
\begin{align}
	{\mathcal{E}} =\left( \begin{array}{ccc}
		c_{\alpha_1} c_{\alpha_2} & s_{\alpha_1} c_{\alpha_2} & s_{\alpha_2}\\
		-(c_{\alpha_1} s_{\alpha_2} s_{\alpha_3} + s_{\alpha_1} c_{\alpha_3})
		& c_{\alpha_1} c_{\alpha_3} - s_{\alpha_1} s_{\alpha_2} s_{\alpha_3}
		& c_{\alpha_2} s_{\alpha_3} \\
		- c_{\alpha_1} s_{\alpha_2} c_{\alpha_3} + s_{\alpha_1} s_{\alpha_3} &
		-(c_{\alpha_1} s_{\alpha_3} + s_{\alpha_1} s_{\alpha_2} c_{\alpha_3})
		& c_{\alpha_2}  c_{\alpha_3}
	\end{array} \right)
	\label{eq:mixingmatrix}
\end{align}
%
where the mixing angles $\alpha_1$, $\alpha_2$ and $\alpha_3$ vary within the ranges :
\begin{eqnarray}
	- \frac{\pi}{2} \le \alpha_{1,2,3} \le \frac{\pi}{2} \;.
\end{eqnarray}
The rotation between the two basis ($\rho_{1},\rho_{2},\rho_{0}$) and ($h_1,h_2,h_3$) diagonalizes the mass matrix ${\mathcal{M}}_{{\mathcal{CP}}_{even}}^2$ as,
\begin{eqnarray}
	{\mathcal{E}}{\mathcal{M}}_{{{\mathcal{CP}}_{even}}}^2{\mathcal{E}}^T&=&diag(m^2_{h_1},m^2_{h_2},m^2_{h_3})
	\label{rota-matrix-cp-even}
\end{eqnarray}
yielding three mass eigenstates, ordered as :
\begin{eqnarray}
	m^2_{h_1} < m^2_{h_2} < m^2_{h_3} 
\end{eqnarray}	
The Yukawa Lagrangian $\mathcal{L}_{\rm Yukawa}$ encompasses the entire Yukawa sector of the $2HDM$ along with an additional Yukawa term originating from the triplet field. The latter generates mass terms for the neutrinos, after spontaneous symmetry breaking.
\begin{equation}
	-\mathcal{L}_{\rm Yukawa} \supset  - Y_{\nu} L^T C \otimes i \sigma^2 \Delta L  + {\rm h.c.} \label{eq:yukawa}
\end{equation}
\noindent
where $L$ and $Y_{\nu}$ denote $SU(2)_L$  doublets of left-handed leptons and neutrino Yukawa couplings, respectively. $C$ is the charge conjugation operator. Furthermore, in terms of the various $\alpha_i$ which appear in the expressions of $\mathcal{E}_{ij}$ matrix elements, we list in Table \ref{table1}, the three ${\mathcal{CP}}_{even}$ Higgs bosons Yukawa couplings for different Yukawa textures.

\begin{table}[t]
	\begin{center}
		\begin{tabular}{|c|c|c|c|c|c|c|c|c|c|c|c|}
			\hline  & $C^{h_1}_U$    & $C^{h_1}_D$ & $C^{h_1}_l$   &   $C^{h_2}_U$   &   $C^{h_2}_D$ &   $C^{h_2}_l$   &   $C^{h_3}_U$  &   $C^{h_3}_D$ &   $C^{h_3}_l$  
			 \\
			\hline  Type-I & $\displaystyle\frac{\mathcal{E}_{12}}{s_\beta} $ & $\displaystyle\frac{\mathcal{E}_{12}}{s_\beta} $&
			$\displaystyle\frac{\mathcal{E}_{12}}{s_\beta} $& $\displaystyle\frac{\mathcal{E}_{22}}{s_\beta} $ & $\displaystyle\frac{\mathcal{E}_{22}}{s_\beta} $ &
			$\displaystyle\frac{\mathcal{E}_{22}}{s_\beta} $ & $\displaystyle\frac{\mathcal{E}_{32}}{s_\beta} $ & $\displaystyle\frac{\mathcal{E}_{32}}{s_\beta} $ &
			$\displaystyle\frac{\mathcal{E}_{32}}{s_\beta} $ 
			\\  
			\hline
		\end{tabular}
	\end{center} 
	\caption{\small
		Normalized Yukawa couplings coefficients of the CP-even $h_i$ neutral Higgs bosons to the leptons, up and down quarks ($u,d$) in 2HDMcT.}
	\label{table1}
\end{table}
On the other hand, by expanding the covariant derivative ${\rm D}_\mu$ and by performing the usual transformations of the
gauge and scalar fields, one can identify the Higgs couplings $h_i$ to the massive gauge bosons $V=W,Z$ as given in Table \ref{table2}, where the two couplings $C^{h_i}_{W^\pm}$ and $C^{h_i}_{Z}$ differ only by a factor 2 associated to $v_t$. Note that in $2HDMcT$ new contributions show up due to the triplet field $\Delta$ coupling  to the SM particles,
\begin{table*}[t]
	\begin{center}
		\setcellgapes{4pt}
		\begin{tabular}{|c|c|c|}
			\hline  
			& $C^{h_i}_W$    & $C^{h_i}_Z$  \\
			\hline  $h_1$ & $\displaystyle{\frac{v_1}{v} \mathcal{E}_{11} + \frac{v_2}{v} \mathcal{E}_{21} + 2\,\frac{v_t}{v} \mathcal{E}_{31}}$ & 
			$\displaystyle{\frac{v_1}{v} \mathcal{E}_{11} + \frac{v_2}{v} \mathcal{E}_{21} + 4\,\frac{v_t}{v} \mathcal{E}_{31}}$  

			  \\
			\hline 
		\end{tabular}
		\caption{The normalized couplings of the neutral ${\mathcal{CP}}_{even}$ $h_1$ Higgs boson to the massive gauge bosons $V=W,Z$ in $2HDMcT$.}
		\label{table2}
	\end{center}
\end{table*}
\section{Constraints and Numerical Results}
\label{section3}
\small
In this section we first present a summary of the theoretical and experimental constraints imposed on the model parameter space \cite{Ouazghour:2018mld,Ouazghour:2023eqr}, then we proceed with an overview of our  analysis and discuss the relevant details of the results, aiming to determine whether the various excesses observed at about near $95$ GeV  can be interpreted as the decay signal of the lighter neutral scalar $h_1$  predicted in  2HDMcT. Here, the  $h_2$ state is chosen to mimic the $125 \GeV$ observed Higgs boson \cite{ATLAS:2012yve,CMS:2012qbp}. Subsequently our analysis will be performed using the type-I Yukawa texture, \footnote{The other Yukawa textures are unable to explain these excesses: In the parameter space of 2HDMcT Types II, III, and IV, the points that satisfy all constraints fall well outside the $2\sigma$ ellipse, and thus cannot account for the observed excesses.}, and the full set of the following theoretical constraints ~\cite{Ouazghour:2018mld,Ouazghour:2023eqr} and Higgs exclusion limits:
\begin{itemize}
	\item \textbf{Unitarity} : The scattering processes must obey the  perturbative unitarity.
	\item \textbf{Perturbativity}: The quartic couplings of the scalar potential are constrained by the following conditions:$| \lambda_i|<8 \pi$ for each $i=1,..,5$.
	\item \textbf{Vacuum stability} : Boundedness from below $BFB$ arising from the positivity in any direction of the fields $\Phi_i$, $\Delta$.
	\item[\textbullet]{\bf Electroweak precision observables}: The oblique parameters $S, T$ and $U$~\cite{Peskin:1991sw,Grimus:2008nb} have been calculated in 2HDMcT~\cite{Ouazghour:2023eqr}. The analysis of the precision electroweak data in light of the new PDG mass of the $W$ boson yields~\cite{ParticleDataGroup:2022pth}:	
	\begin{equation}
		S^{exp} = -0.01\pm 0.07,\;\;\;T^{exp} = 0.04\pm 0.06\;\;\;  \;\;\; \rho_{ST} = +0.92.
	\end{equation} 
	We use the following $\chi^2_{ST}$ test :
	\begin{eqnarray}
		\chi^2_{ST} &=& \frac{1}{\hat{\sigma_T}^2_{1}(1-\rho_{ST}^2)}(T - T^{exp})^2
		+ \frac{1}{\hat{\sigma_S}^2_{1}(1-\rho_{ST}^2)}(S - S^{exp})^2 \nonumber \\ 
		& -& \frac{2\rho_{ST}}{\hat{\sigma}_{T}\hat{\sigma}_{S}(1-\rho_{ST}^2)}(T - T^{exp})(S - S^{exp})\le R^2\,, \label{eq:chi2}
	\end{eqnarray}
with $R^2=2.3$ and $5.99$ corresponding to $68.3 \%$  and
$95 \% $  confidence levels (C.L.) respectively.
Our numerical analysis is performed with $\chi^2_{ST}$ at 95\% C.L. 
	\item To further delimit the allowed parameter space, the \texttt{HiggsTools} package~\cite{Bahl:2022igd} is employed. This ensures that the allowed parameter regions align with the observed properties of the $125$~GeV Higgs boson  ( \texttt{HiggsSignals}~\cite{Bechtle:2013xfa,Bechtle:2014ewa,Bechtle:2020uwn,Bahl:2022igd}) and with the exclusion limits from additional Higgs bosons search at the LHC and at LEP ( \texttt{HiggsBounds}~\cite{Bechtle:2008jh,Bechtle:2011sb,Bechtle:2013wla,Bechtle:2020pkv,Bahl:2022igd}).
	\item[\textbullet]{\bf Flavour constraints}:  Flavour constraints are also implemented in our analysis by using the $B$-physics results derived in \cite{Ouazghour:2023eqr} and the experimental data at 2$\sigma$ \cite{HFLAV:2022pwe} shown in Table \ref{table3}.
\end{itemize}
{\renewcommand{\arraystretch}{1.5} 
	{\setlength{\tabcolsep}{0.1cm} 
		\begin{table*}[t]
			\centering
			\setlength{\tabcolsep}{7pt}
			\renewcommand{\arraystretch}{1.2} %
			\begin{tabular}{|l||c|c|}
				\hline
				Observable & Experimental result & 95\% C.L.\\\hline
				BR($\bar{B}\to X_{s}\gamma$) \cite{Ouazghour:2023eqr}&$(3.49\pm 0.19)\times10^{-4}$ \cite{HFLAV:2022pwe}&$[3.11\times 10^{-4} , 3.87\times 10^{-4}]$\\\hline
			\end{tabular}
			\caption{Experimental result of flavor observable: $\bar{B}\to X_{s}\gamma$ at 95$\%$ C.L.}
			\label{table3}
		\end{table*}
		Our analysis is performed using the following input parameters in 2HDMcT,
\begin{align}
	\mathcal{P}_I=\left\{\alpha_1,\alpha_2,\alpha_3,m_{h_1},m_{h_2}, \lambda_{1},\lambda_{3},\lambda_{4},\lambda_{6},\lambda_{7},\lambda_{8},\lambda_{9},\bar{\lambda}_{8},\bar{\lambda}_{9}\right. 
	\left.,\mu_1,v_t,\tan\beta \right\}
	\label{eq:set-para1}
\end{align}

		in the following ranges,
		\begin{align}
			\begin{matrix}
				93\,\text{GeV}\leq m_{h_1}\leq 97\,\text{GeV},\,\,m_{h_1}\leq m_{h_2}\leq m_{h_3}\leq 1\,\text{TeV},\,\,
				\,\,80\,\text{GeV}\leq m_{A_1}\leq m_{A_2}\leq 1\,\text{TeV}\,\\
				80\,\text{GeV}\leq m_{H^{\pm\mp}}\leq 1\,\text{TeV},\,80\,\text{GeV}\leq m_{H_1^{\pm}}\leq m_{H_2^{\pm}}\leq 1\,\text{TeV},\,\,
				m_{h_2}=125.09\,\,\text{GeV},\,\, \frac{-\pi}{2}\leq \alpha_{1}\leq \frac{\pi}{2},\,\, -0.1\leq \alpha_{2,3}\leq 0.1\\
				0.5\leq \tan\beta=v_2/v_1\leq 30,\,\,-10^2\leq \mu_1\leq 10^2,\,\,0\leq v_t\leq 1 \,\text{GeV}\,\,\, \,\,-8\pi\leq \lambda_{i},\bar{\lambda_{i}}\leq 8\pi,
			\end{matrix}
			\label{input}
		\end{align}
		
		We assess the agreement with the $95$ GeV excesses using a $\chi^2$-test defined as,
		\begin{eqnarray}
			\chi^2_{\gamma\gamma,b\bar{b},\tau^+\tau^-} = \frac{(\mu_{\gamma\gamma,b\bar{b},\tau^+\tau^-} -\mu_{\gamma\gamma,b\bar{b},\tau^+\tau^-}^{\rm exp})^2 }{(\Delta \mu_{\gamma\gamma,b\bar{b},\tau^+\tau^-}^{\rm exp})^2} \ ,
			\label{eqmu}
		\end{eqnarray}
		where $\mu_{\rm \gamma\gamma}$, $\mu_{\mathrm{b\bar{b}}}$ and $\mu_{\rm \tau^+\tau^-}$ are signal strengths defined in terms of the production cross sections  and decay branching ratios as follows,
		
		\begin{eqnarray}
			\mu_{\rm \gamma\gamma}&=\frac{\sigma_{\rm 2HDMcT}(gg\to h_1)}{\sigma_{\rm SM}(gg\to h_{\rm SM})}\times \frac{{\cal BR}_{\rm 2HDMcT}(h_1\to \gamma\gamma)}{{\cal BR}_{}(h_{\rm SM}\to \gamma\gamma)}
			\approx\left|c_{h_1t\bar{t}}\right|^2\times \frac{{\cal BR}_{\rm 2HDMcT}(h_1\to \gamma\gamma)}{{\cal BR}_{}(h_{\rm SM}\to \gamma\gamma)},\\\nonumber\\
			\mu_{\mathrm{b\bar{b}}} &= \frac{\sigma_{\rm 2HDMcT}(e^+e^-\to Zh_1)}{\sigma_{\rm SM}(e^+e^-\to Zh_{\rm SM})}\times \frac{{\cal BR}_{\rm 2HDMcT}(h_1\to b\bar{b})}{{\cal BR}_{}(h_{\rm SM}\to b\bar{b})} \approx \left|c_{h_1ZZ}\right|^2\times \frac{{\cal BR}_{\rm 2HDMcT}(h_1\to b\bar{b})}{{\cal BR}_{}(h_{\rm SM}\to b\bar{b})},\\\nonumber\\
			\mu_{\rm \tau^+\tau^-}&=\frac{\sigma_{\rm 2HDMcT}(gg\to h_1)}{\sigma_{\rm SM}(gg\to h_{\rm SM})}\times \frac{{\cal BR}_{\rm 2HDMcT}(h_1\to \tau^+\tau^-)}{{\cal BR}_{}(h_{\rm SM}\to \tau^+\tau^-)}
			\approx\left|c_{h_1t\bar{t}}\right|^2\times \frac{{\cal BR}_{\rm 2HDMcT}(h_1\to \tau^+\tau^-)}{{\cal BR}_{}(h_{\rm SM}\to \tau^+\tau^-)}
			\label{coup1}
		\end{eqnarray} 	
		
		\noindent
		The experimental values for $\mu_{\gamma\gamma,b\bar{b},\tau^+\tau^-}^{\rm exp} $ reported by CMS \cite{CMS-PAS-HIG-20-002,CMS:2024yhz} and ATLAS \cite{ATLAS-CONF-2023-035,ATLAS:2024bjr,Biekotter:2023oen} are given by \footnote{The value of $\mu_{\gamma \gamma}^\text{ATLAS}$ is found by normalizing the cross section reported by ATLAS \cite{ATLAS:2024bjr} to its corresponding SM one \cite{LHCHiggsCrossSectionWorkingGroup:2016ypw}.}, 
		\begin{equation}
			\mu_{\gamma \gamma}^\text{CMS}
			= 0.33^{+0.19}_{-0.12} 
		\end{equation}
		\begin{equation}
			\mu_{\gamma \gamma}^\text{ATLAS}
			= 0.18^{+0.10}_{-0.10} 
		\end{equation}
		Here $\sigma_{\rm SM}$ denotes the cross section of a hypothetical SM Higgs $h_{\rm SM}$ at the same mass.  
		
		\noindent
		An earlier analysis combined the two results in a way overlooking potential correlations. This analysis suggests a $3.1\sigma$ (local) excess at a mass of $95.4$ GeV, with a signal strength \cite{Biekotter:2023oen},
		\begin{align}
			\mu_{\gamma \gamma}^\text{ATLAS+CMS} = 0.24^{+0.09}_{-0.08}.
		\end{align}
		The CMS experiment also observed a similar excess in the $\tau^+\tau^-$ channel  with $\mu_{\tau^+\tau^-}^{CMS} = 1.2 \pm 0.5$, corresponding to a local significance $2.6 \sigma$ \cite{CMS:2022goy}.
		
		Also, it is worth noting that LEP reported a local excess of $2.3 \sigma$ in the $b \bar{b}$  channel \cite{LEPWorkingGroupforHiggsbosonsearches:2003ing} with a signal strength, 
		\begin{equation}
			\mu_{b\bar{b}}^{LEP} = 0.117 \pm 0.057.
		\end{equation}

		\subsection{ $h_{95}\to\gamma\gamma$ and $h_{95}\to b\bar{b}$ excesses}
		\label{section4}
		In this section, we examine whether the 2HDMcT allowed parameter space can simultaneously account for the $95$ GeV excesses observed in the two decay modes: $h_1\to \gamma\gamma, b\bar{b}$.
		To this end we perform a $\chi^2$ -test which incorporates the two contributions $\chi_{\gamma\gamma}^2$ and $\chi_{b\bar{b}}^2$,  knowing that the latter  depend on the model predictions for $\mu_{\gamma\gamma}$ and $\mu_{b\bar{b}}$ (see Eq.\ref{eqmu}). In order to insure that the SM-like Higgs boson $h_2$ aligns with the observed properties of the $125$~GeV Higgs  $h_{125}$, we also use $\chi_{125}^2$ and impose the following condition $\Delta{\chi_{125}^2}<5.99$\footnote{The $\chi_{125}^2$ is provided by \texttt{HiggsSignals}. The  $\Delta \chi_{125}^2 $ condition is defined as, $\Delta \chi_{125}^2 = \chi^2 - \chi^2_{\mathrm{min}}$} \cite{Bahl:2022igd}. Here we impose $\Delta \chi_{125}^2 < 5.99$, with:
		\begin{equation}
			\chi^2_{\gamma\gamma+b\bar{b}} =\chi^2_{\gamma\gamma}+\chi^2_{b\bar{b}}
			\label{eq:yeslep}
		\end{equation}
	and
		\begin{equation}
			\chi^2_{Tot}=\chi^2_{\gamma\gamma}+\chi^2_{b\bar{b}}+\chi^2_{125}
			\label{chi-total}
		\end{equation}
		In Fig. \ref{fig1}, we show the scatter points of the scan results in ($\mu_{\gamma\gamma}, \mu_{b\bar{b}}$) plane within 2HDMcT. The color bar shows the couplings $C_{h_1 ZZ}$ (left panel) and $C_{h_1 t\bar{t}}$ (right panel) values. The regions consistent with the excesses at $1\sigma$ are represented by the dashed ellipses defined by the equation $\chi^2_{b\bar{b}} + \chi^2_{\gamma\gamma} = 2.30$. The black, green, and blue ellipses correspond to the $\chi^2$ evaluated via the experimental signal strengths $\mu_{\gamma\gamma}^{\mathrm{CMS+ATLAS}}$, $\mu_{\gamma\gamma}^{\mathrm{ATLAS}}$, and $\mu_{\gamma\gamma}^{\mathrm{CMS}}$, respectively. The green, red and orange stars fingerprint the points with $min(\chi^2_{\gamma\gamma+b\bar{b}}$), $min(\chi^2_{125}$) and the best fit point\footnote{It is worth to recall that the best-fit point corresponds to the minimum value of $\chi^2_{Tot}$.}, respectively. The set of generated points passing  $\Delta{\chi_{125}^2}<5.99$ condition are depicted in black. We can clearly see that the 2HDMcT with Type-I Yukawa texture is capable of accounting for the two excesses simultaneously as a dense set of  points occur within the $1\sigma$ ellipse. Furthermore, we also see that higher values of the couplings $C_{h_1 ZZ}$ and $C_{h_1 t\bar{t}}$ correspond to larger 
		signal strengths $\mu_{\gamma\gamma}$, $\mu_{b\bar{b}}$, as expected from  Eq. \ref{coup1}.
		
		\begin{figure}[hbtp]
			\centering
			\includegraphics[width=0.83\textwidth]{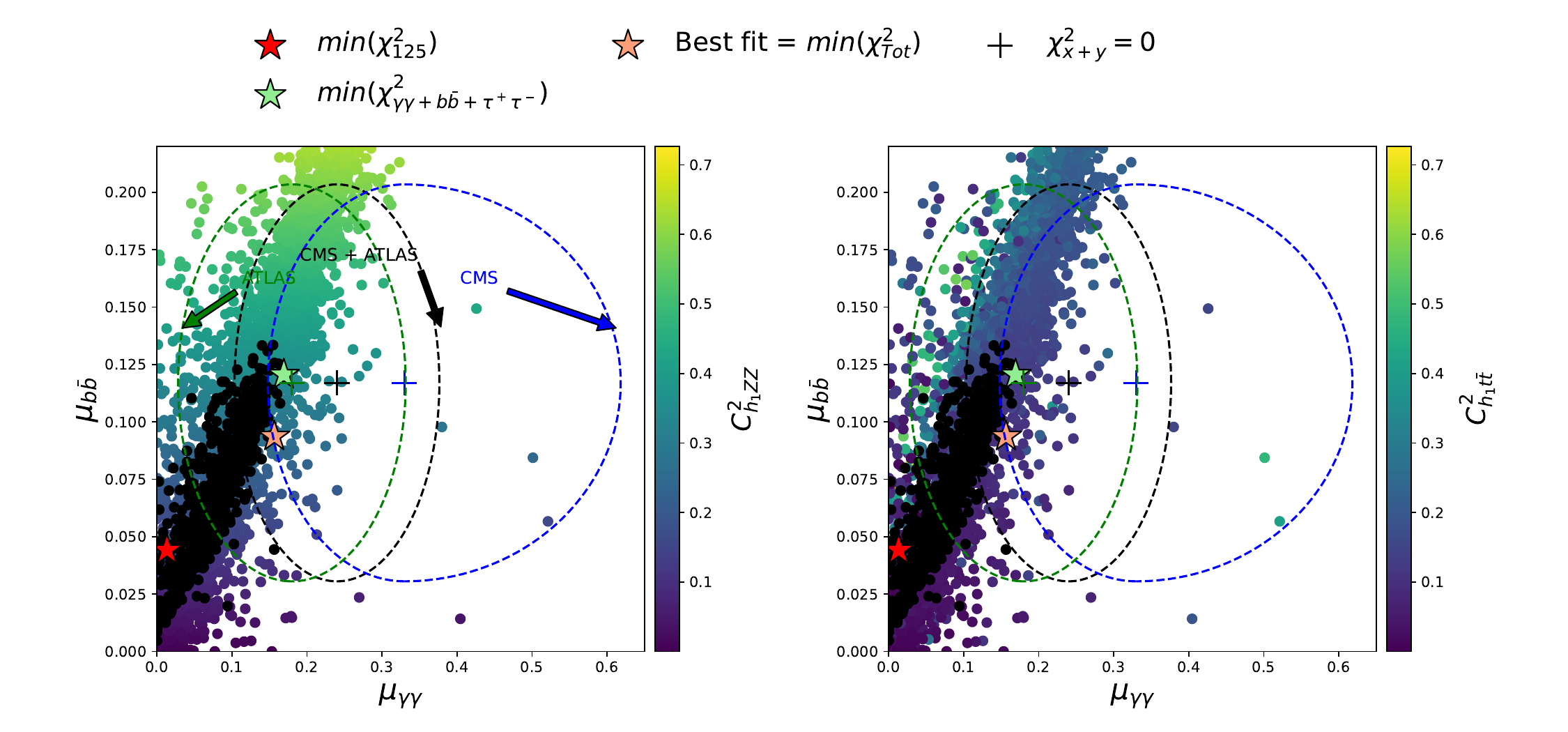}~

			\caption{\small 
				The scatter points in ($\mu_{b\bar{b}}$, $\mu_{\gamma\gamma}$) plane in the 2HDMcT. The color coding indicates the values of the couplings $C_{h_1 ZZ}$ (left panel) and $C_{h_1 t\bar{t}}$ (right panel). Confidence levels of $1\sigma$ with respect to $\chi^2_{\gamma\gamma+b\bar{b}}$ are illustrated by dashed ellipses. Black, green and blue represent the predictions for $\mu_{\gamma\gamma}$ using the CMS+ATLAS, ATLAS and CMS data, respectively. Only the points in black satisfy the condition  $\Delta{\chi_{125}^2}<5.99$. The green, red and orange stars indicate the points with $min(\chi^2_{\gamma\gamma+b\bar{b}}$), $min(\chi^2_{125}$) and the best fit point, respectively.}
			\label{fig1}
		\end{figure}
		\begin{figure}[hbtp]
			\centering
			\includegraphics[width=1.1\textwidth]{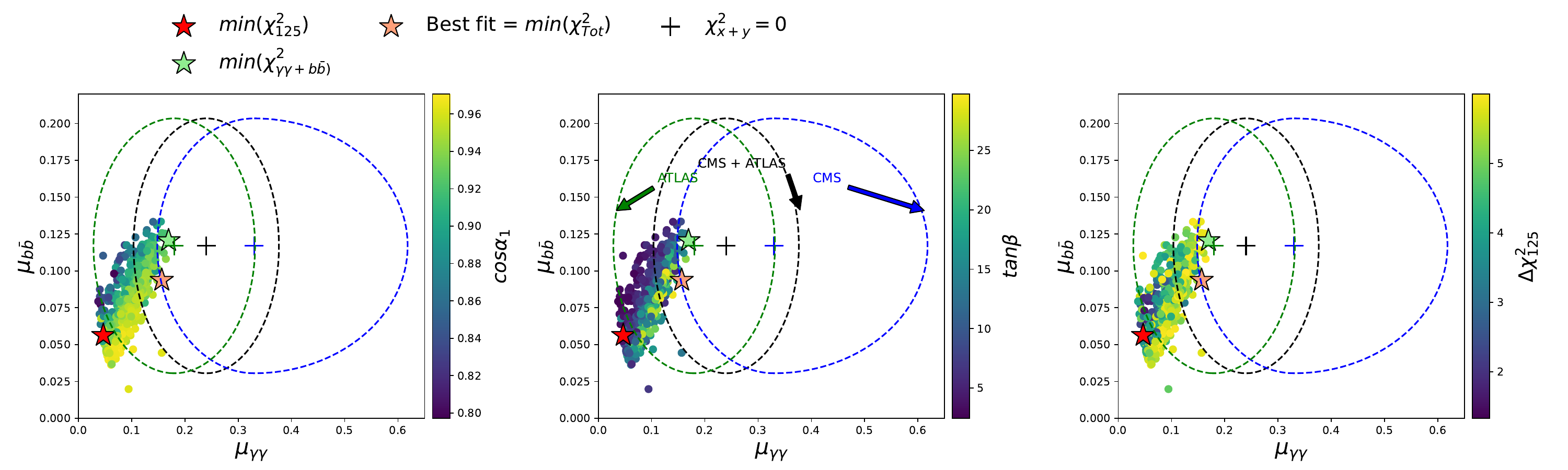}~
			\caption{\small As in Fig.\ref{fig1} the scatter points in ($\mu_{\gamma\gamma}$, $\mu_{b\bar{b}}$) plane in the 2HDMcT are presented. The color coding indicates the values of cos $\alpha_1$ (left), tan $\beta$ (middle) and $\Delta{\chi_{125}^2}$ (right). The green, red and orange stars marking the points with $min(\chi^2_{\gamma\gamma+b\bar{b}}$), $min(\chi^2_{125}$) and the best fit point, respectively.}
			\label{fig2}
		\end{figure}
		Similarly Fig. \ref{fig2} displays scattered points in the ($\mu_{\gamma\gamma}$, $\mu_{b\bar{b}}$) plane. The color bar displays the values of  $ \cos\alpha_1$ (left),  $tan \beta$ (middle), and $\Delta{\chi_{125}^2}$ (right). For a given $\mu_{b\bar{b}}$, the largest $\mu_{\gamma\gamma}$  corresponds to largest  $tan\beta$. In contrast, for a fixed $\mu_{\gamma\gamma}$,  we get sizable $\mu_{b\bar{b}}$ when $\tan \beta$ is smaller. A similar pattern occurs with $ \cos \alpha_1$, except that most allowed points originate from relatively large values of $ \cos \alpha_1$. Additionally, we can also see that $\mu_{\gamma\gamma}$ increases when $\Delta{\chi_{125}^2}$ gets large,  corresponding to generating points representing most accurately the excesses. Consequently, the point with $min(\chi^2_{125})$, signaled by a red star, lies outside the $1\sigma$ ellipse. In order to probe how the excesses affects the parameters $\cos \alpha_{1}$ and $\ tan \beta$,  we display in Fig. \ref{chi2_1} $\chi^2_{\gamma\gamma+b\bar{b}}$ versus $\cos \alpha_1$, where the color coding indicates $tan \ \beta$ values. We can readily deduce that simultaneously describing both excesses at the $1 \sigma$ confidence level requires these parameters to lie within the following intervals :
		\begin{equation}
			0.83\leq cos\ \alpha_{1}\leq 0.96,\,\, 4\leq \tan\beta\leq 30
		\end{equation}
		\begin{figure}[hbtp]
	\centering
	\includegraphics[width=0.40\textwidth]{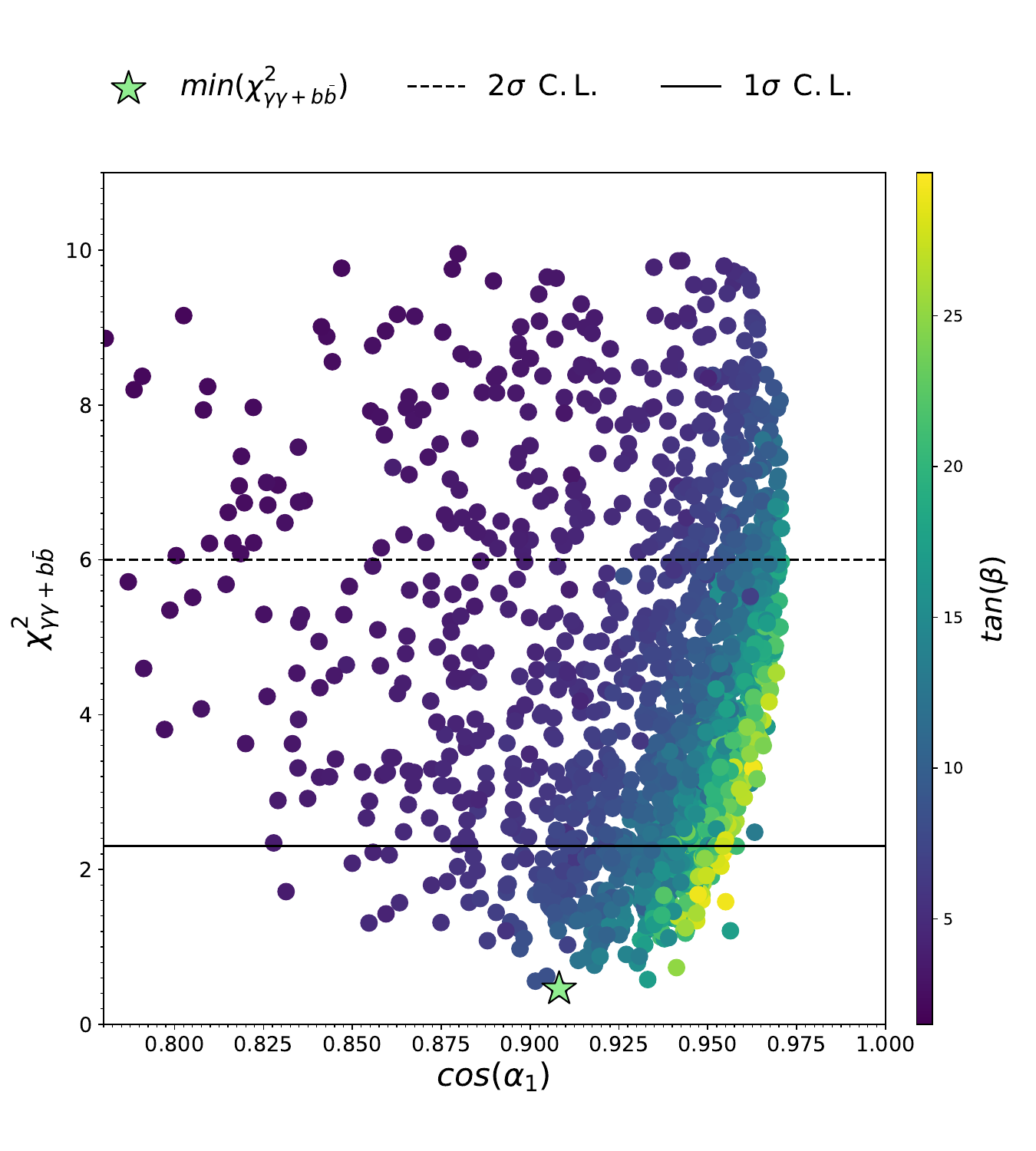}~
	\caption{\small
		Scattered points in the plane ($\cos \alpha_1$, $\ \chi^2_{\gamma\gamma+b\bar{b}}$). The color bar indicates the value of $\tan\beta$. The horizontal solid and dashed lines represent the $1\sigma$ and $2\sigma$ regions, respectively. The green star indicates the point with $min(\chi^2_{\gamma\gamma+b\bar{b}})$.
	}
	\label{chi2_1}
\end{figure}
		\subsection{ $h_{95}\to\gamma\gamma$, $h_{95}\to b\bar{b}$  and $h_{95}\to \tau^+\tau^-$ excesses}
		\label{section5}
		Besides the $\gamma\gamma$ excess, the CMS experiment has also reported an excess in the $\tau^+\tau^-$ decay channel with a local significance of $2.6 \sigma$ \cite{CMS:2022goy}. Here, we explore whether the 2HDMcT model is able to provide a unified explanation to the three observed excesses. To do so, we also incorporate the contribution from $\chi_{\tau^+\tau^-}^2$ and define the following $\chi^2$ tests:
		\begin{equation}
			\chi^2_{\gamma\gamma+b\bar{b}+\tau^+\tau^-} =\chi^2_{\gamma\gamma}+\chi^2_{b\bar{b}}+\chi^2_{\tau^+\tau^-}
			\label{eq:yeslep2}
		\end{equation}
		and 
		\begin{equation}
			\chi^2_{Tot} =\chi^2_{\gamma\gamma}+\chi^2_{b\bar{b}}+\chi^2_{\tau^+\tau^-}+\chi^2_{125}
			\label{chi-total2}
		\end{equation}
			\begin{figure*}[hbtp]
		\centering
		\includegraphics[width=1.08\textwidth]{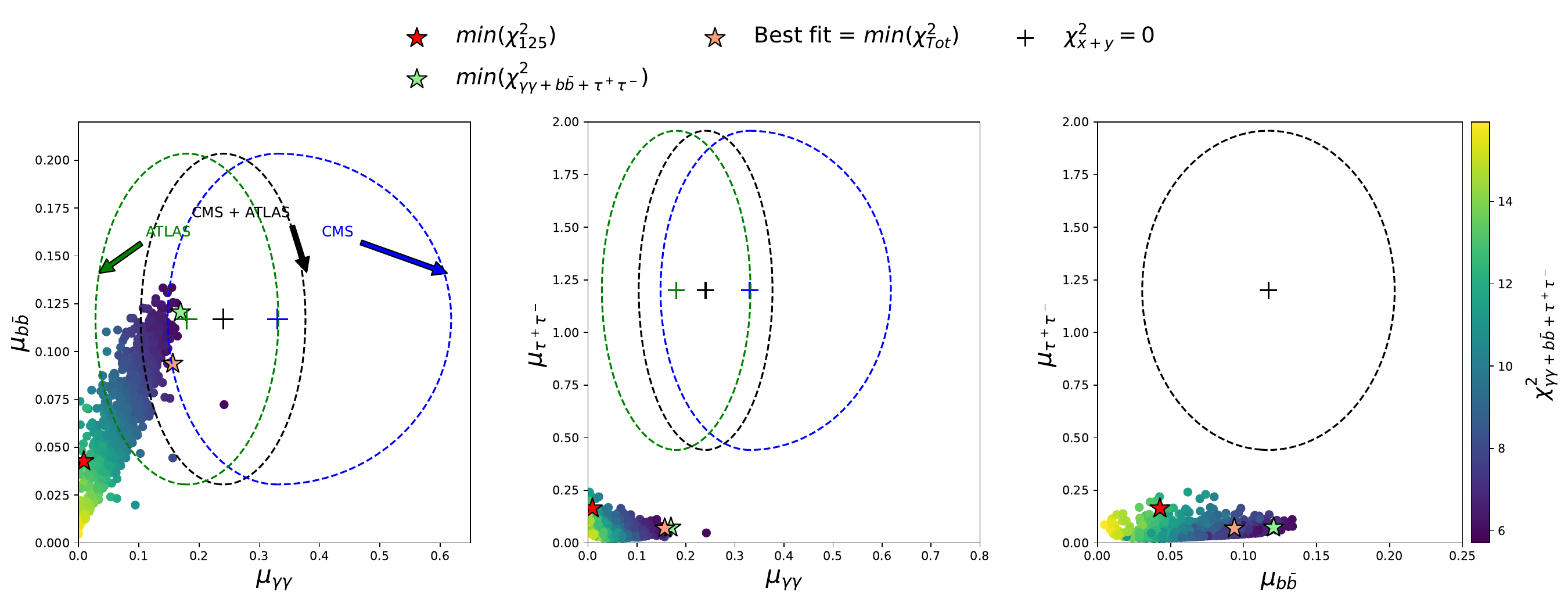}~
		\caption{\small The colour map of $\chi^2_{\gamma\gamma+\tau^{+}\tau^{-}+b\bar{b}}$ in the 
			($\mu_{\gamma\gamma}, \mu_{b\bar b}$), ($\mu_{\gamma\gamma}, \mu_{\tau^{+}\tau^{-}})$ 
			and ($\mu_{b\bar b}-\mu_{\tau^{+}\tau^{-}}$)
			planes of the 
			signal strength parameters for  2HDMcT parameter space under study.
			The ellipses define the
			regions consistent with the excess at 1$\sigma$ C.L.
			The black, green, and blue ellipses correspond to the $\chi^2$ values calculated using the experimental signal strengths $\mu_{\gamma\gamma}^{\mathrm{CMS+ATLAS}}$, $\mu_{\gamma\gamma}^{\mathrm{ATLAS}}$, and $\mu_{\gamma\gamma}^{\mathrm{CMS}}$, respectively. The green, red and orange stars marking the points with $min(\chi^2_{\gamma\gamma+b\bar{b}+\tau^{+}\tau^{-}}$), $min(\chi^2_{125}$) and the best fit point, respectively.}
		\label{fig3}
	\end{figure*}
        In Fig.~\ref{fig3}, the scattered points in the  signal strengths ($\mu_{\gamma\gamma}$, $\mu_{b \bar b}$) (left), ($\mu_{\gamma\gamma}$, $\mu_{\tau^{+}\tau^{-}}$) (middle) and ($\mu_{b \bar b}$, $\mu_{\tau^{+}\tau^{-}}$) (right)  planes are presented. The dashed ellipses indicate regions consistent with the two-dimensional 1$\sigma$ excess in ($\mu_{x}, \mu_{y}$), described by the equation $\chi^2_x + \chi^2_y = 2.30$, where the subscripts $x$ and $y$ represent each possible pairing of the three signal channels ($\gamma\gamma$, $\tau^{+}\tau^{-}$, and $b\bar{b}$). The black, green, and blue ellipses show the $\chi^2$ values computed from the experimental signal strengths $\mu_{\gamma\gamma}^{\mathrm{CMS+ATLAS}}$, $\mu_{\gamma\gamma}^{\mathrm{ATLAS}}$, and $\mu_{\gamma\gamma}^{\mathrm{CMS}}$, respectively. The green, red and orange stars marking the points with $min(\chi^2_{\gamma\gamma+b\bar{b}}$), $min(\chi^2_{125}$) and the best fit point, respectively. The color coding indicates the value of $\chi^2_{\gamma\gamma+\tau^{+}\tau^{-}+b\bar{b}}$. We can clearly see from the left panel that many points lies solely within the 1$\sigma$ C.L. ellipse for the ($\mu_{b\bar{b}}$, $\mu_{\gamma\gamma}$) pair, however no point do so for the other two pairs since it is quite difficult to achieve larger values for the two pairs simultaneously. Consequently the model cannot  account for all three excesses at 1 $\sigma$ C.L simultaneously. The point indicated by the green star corresponding to $min(\chi^2_{\gamma\gamma+\tau^+\tau^-+b\bar{b}})$ has a value of $5.7$, corresponding to about $1.4 \sigma$ C.L. for three degrees of freedom. So by assuming that the observed signal originates only from the CP-even Higgs $h_1$, the 2HDMcT can explain the observed excess across all three channels simultaneously at about 1.4 $\sigma$ C.L. at best  
        
        Similarly in Fig. \ref{fig4}, we plotted the allowed points in ($\mu_{\gamma\gamma}$, $\mu_{b\bar{b}}$) (upper panels) and ($\mu_{\gamma\gamma}$, $\mu_{\tau^+\tau^-}$),($\mu_{b\bar{b}}$, $\mu_{\tau^+\tau^-}$) (lower panels) planes. The color coding indicates the values of cos $\alpha_1$ (left panels) and tan $\beta$ (right panel). In the upper panels of Fig. \ref{fig4}, we observe a similar dependence of $\mu_{\gamma\gamma}$ and $\mu_{b\bar{b}}$ on $\tan \beta$ and $\cos \alpha_1$ as seen in Fig. \ref{fig2}. Specifically, $\mu_{\gamma\gamma}$ gets large  for sizable values of $\tan \beta$ and $\cos \alpha_1$. In contrast, the lower panels of Fig. \ref{fig4} show that the points with smallest $\tan \beta$ and $\cos \alpha_1$ produce larger $\mu_{\tau^+\tau^-}$. Consequently, our analysis show that it is not possible to achieve large values for both $\mu_{\gamma\gamma}$ and $\mu_{\tau^+\tau^-}$ simultaneously, thereby preventing the simultaneous $1\sigma$ accounting for both excesses.
		\begin{figure*}[t]
			\centering
			\includegraphics[width=0.8\textwidth]{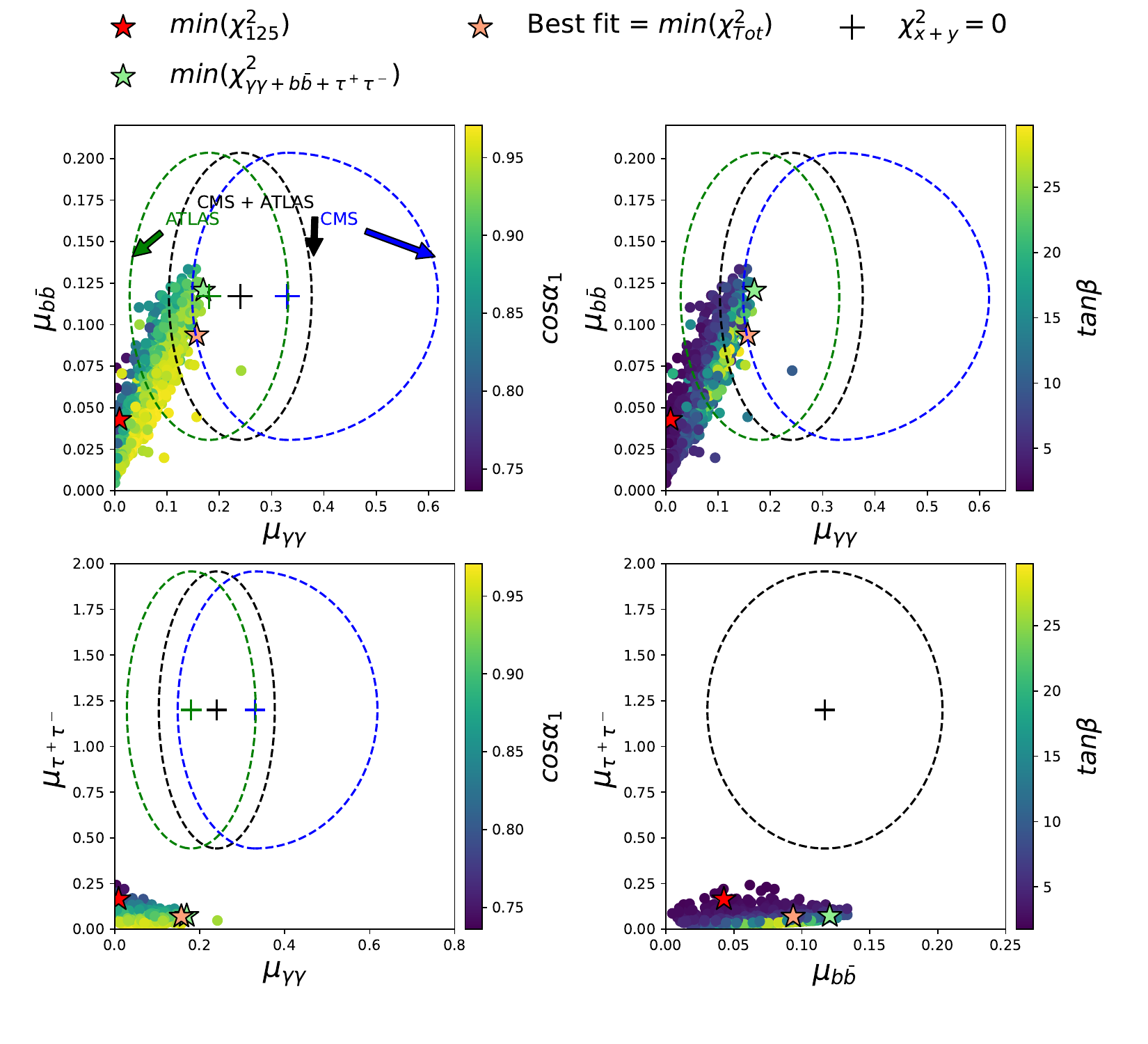}~

			\caption{\small Just as in Fig. \ref{fig3} the scatter points in ($\mu_{\gamma\gamma}$, $\mu_{b\bar{b}}$) (upper panels) and ($\mu_{\gamma\gamma}$, $\mu_{\tau^+\tau^-}$),($\mu_{b\bar{b}}$, $\mu_{\tau^+\tau^-}$) (lower panels) planes in the 2HDMcT are presented. The color coding indicates the values of cos $\alpha_1$ (left), tan $\beta$ (right). The green, red and orange stars marking the points with $min(\chi^2_{\gamma\gamma+b\bar{b}+\tau^{+}\tau^{-}}$), $min(\chi^2_{125}$) and the best fit point, respectively.}
			\label{fig4}
		\end{figure*}
	In the subsequent section, we will investigate whether a superposition of the CP-even and CP-odd states with a nearly degenerate mass can account for all the observed excesses. 	
		\subsection{The Superposition Solution}
		In the previous sections we have shown that the 2HDMcT can well describe the $\gamma\gamma$ and $b\bar{b}$ excesses simultaneously. On the other hand, we have seen that the model fails to account for the three excesses simultaneously, namely, $\gamma\gamma$, $b\bar{b}$ and $\tau^+\tau^-$ at $1\sigma$ C.L. A legitimate question arise, whether these excesses are a manifest of the superposition of two or more resonances at the comparable mass.
		
		Here, we will explore this scenario and see if the model can explain all the excesses simultaneously. To do this,  we conduct a combined analysis of the CP-even ($h_1$) and CP-odd ($A_1$) resonances\footnote{The input parameters are the same as in \ref{input} plus $93\,\text{GeV}\leq m_{A_1} \leq 97\,\text{GeV}$.}. In the CP-conserving scenario the pseudoscalar state $A_1$ cannot contribute to $\mu_{b\bar{b}}$ given that the coupling $C_{A_1ZZ}$ is forbidden at tree level. On the other hand the state $A_1$ do contribute to both $\mu_{\gamma\gamma}$ and $\mu_{\tau^+\tau^-}$.
		In this case the combined contributions to the signal strengths for the three resonances \(\gamma\gamma\), $b\bar{b}$ and \(\tau^+\tau^-\) , reads as follows:
		\begin{eqnarray}
			\mu_{\gamma\gamma}(h_1+A_1) &=& \mu_{\gamma\gamma}(h_1) + \mu_{\gamma\gamma}(A_1), \nonumber \\
			\mu_{\tau^+\tau^-}(h_1+A_1) &=& \mu_{\tau^+\tau^-}(h_1) + \mu_{\tau^+\tau^-}(A_1),\nonumber \\
			\mu_{b\bar{b}}(h_1+A_1) &=& \mu_{b\bar{b}}(h_1),
			\label{28}
		\end{eqnarray}\footnote{$\mu_{b\bar{b}}(A_1)=0$.}
		
		Since we assumed CP conservation in the 2HDMcT model, there is no interference between the $h_1$ and $A_1$ states. In a similar pattern to Fig. \ref{fig3}, and taking contributions from both states $h_1$ and $A_1$ as defined in Eq. \ref{28}, we display the scattered points in the signal strengths ($\mu_{\gamma\gamma}$, $\mu_{b \bar b}$) (left), ($\mu_{\gamma\gamma}$, $\mu_{\tau^{+}\tau^{-}}$) (middle) and ($\mu_{b \bar b}$, $\mu_{\tau^{+}\tau^{-}}$) (right) planes. We clearly see that a dense set of points lies within the $1 \sigma$ level in both ($\mu_{\gamma\gamma}$, $\mu_{b \bar b}$), ($\mu_{\gamma\gamma}$, $\mu_{\tau^{+}\tau^{-}}$) planes and a few in the ($\mu_{b \bar b}$, $\mu_{\tau^{+}\tau^{-}}$) plane, but in all cases the point with $min(\chi^2_{\gamma\gamma+b\bar{b}+\tau^{+}\tau^{-}}$), indicated by the green star lies within the $1 \sigma$ ellipse. This proves that the 2HDMcT is able to account for the observed excesses at $1\sigma$ C.L when we consider the superposition of  ($h_1+A_1$) states. The $min(\chi^2_{\gamma\gamma+\tau^+\tau^-+b\bar{b}})=2.25$, corresponding to about $0.64 \sigma$ C.L, is the model's best description of the three-dimensional excesses. This point is displayed in Table \ref{description}. The generated points explaining the three excesses at $1\sigma$ ($\chi^2 \leq 3.53$) are marked in black in the figures.
		
		For a direct comparison with the experimental data, we superimpose our predictions for $\mu_{\gamma\gamma}$ onto the CMS $13$ TeV low-mass $\gamma\gamma$ analysis data in Fig.~\ref{mugg}. The expected and observed cross-section limits obtained by CMS (black) and ATLAS (red) are displayed by dashed and solid lines, respectively. The green and yellow bands correspond to the $1\sigma$ and $2\sigma$ uncertainty intervals, respectively. The error bar in yellow indicates the value of $\mu_{\gamma\gamma}^{CMS+ATLAS}$, along with its respective uncertainty. One can readily see that the 2HDMcT can well accommodate the combined  $\mu_{\gamma\gamma}$ observed excess.
			\begin{figure}[htp!]	
		\centering
		\mbox{\includegraphics[height=7cm,width=17cm]{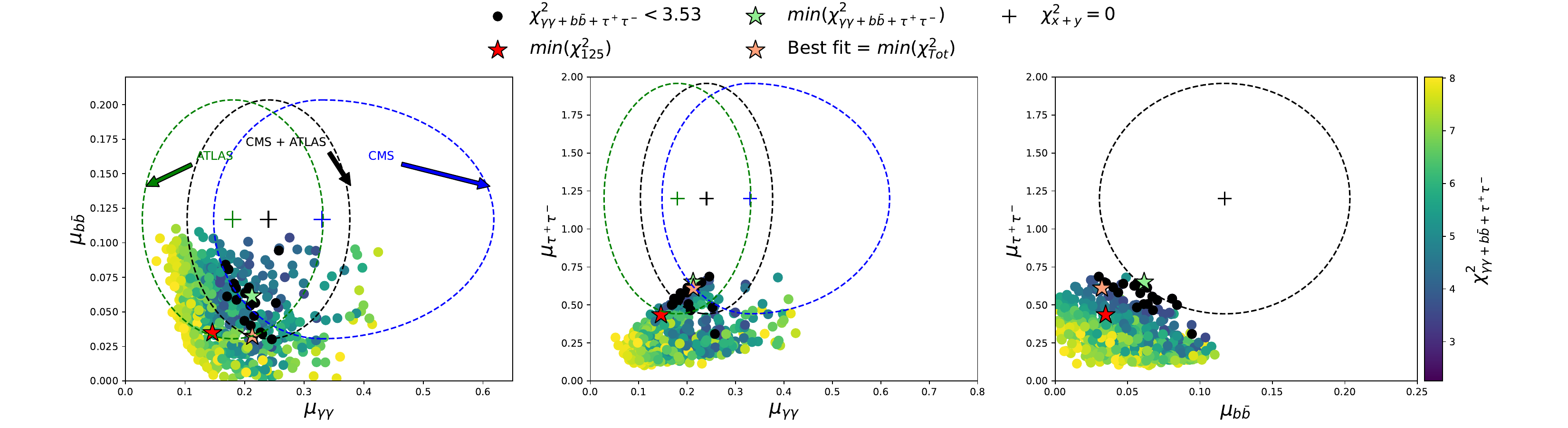}}
		\caption{The scattered points in the plane of ($\mu(h_1+A_1)_{\gamma\gamma}$, $\mu(h_1+A_1)_{\tau^+\tau^-}$) within the 2HDMcT. The green, red and orange stars marking the points with $min(\chi^2_{\gamma\gamma+b\bar{b}+\tau^+\tau^-}$), $min(\chi^2_{125}$) and the best fit point, respectively. The black points feature $\chi^2_{\gamma\gamma+\tau^+\tau^-+b\bar{b}} \leq 3.53$
			and describe the excesses at the
			level of $1\,\sigma$ or better. }
		\label{fig7}
	\end{figure}

		In order to see how the excesses impact the parameters $\cos\alpha_1$ and $\tan \beta$ and compare with the results shown in Fig. \ref{chi2_1}, we display in Fig. \ref{chi2}, $\chi^2_{\gamma\gamma+\tau^+\tau^-+b\bar{b}}$ with  parameter scans as a function of $cos\ \alpha_1$. The color bar indicates $tan \ \beta$ values. The horizontal solid and dashed lines represent the $1\sigma$ and $2\sigma$ regions respectively. We can clearly see that describing simultaneously the three excesses at $1\sigma$ C.L is possible, imposing constraints on the allowed ranges of $\tan\beta$ and $cos\alpha_1$. In particular,  these parameters lie within a drastically reduced intervals compared to those found in Sec. \ref{section4}, namely :
		\begin{equation}
			0.78\leq cos\ \alpha_{1}\leq 0.86,\,\, 2\leq \tan\beta\leq 5
		\end{equation}
				\begin{figure*}[hbtp]
			\centering
			\includegraphics[width=0.350\textwidth]{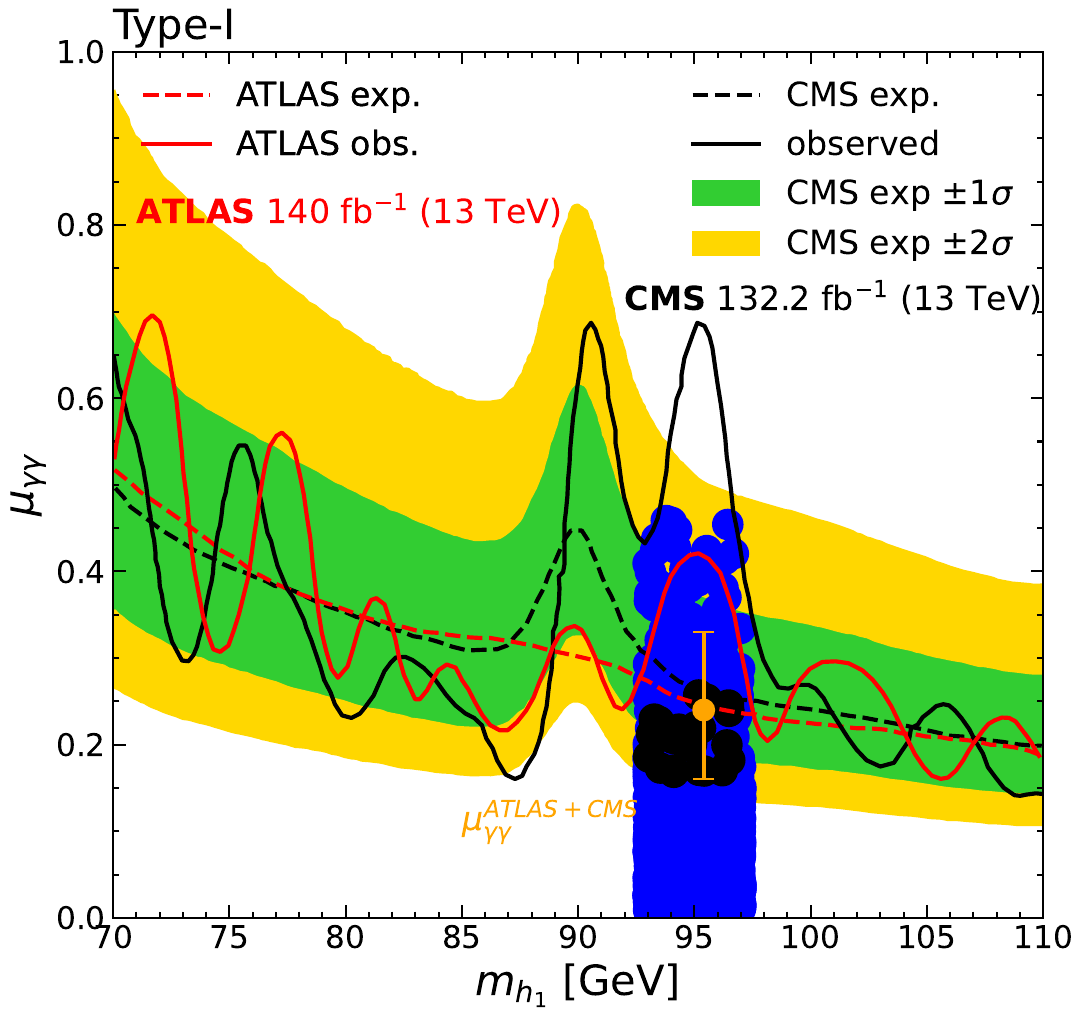}~
			\includegraphics[width=0.350\textwidth]{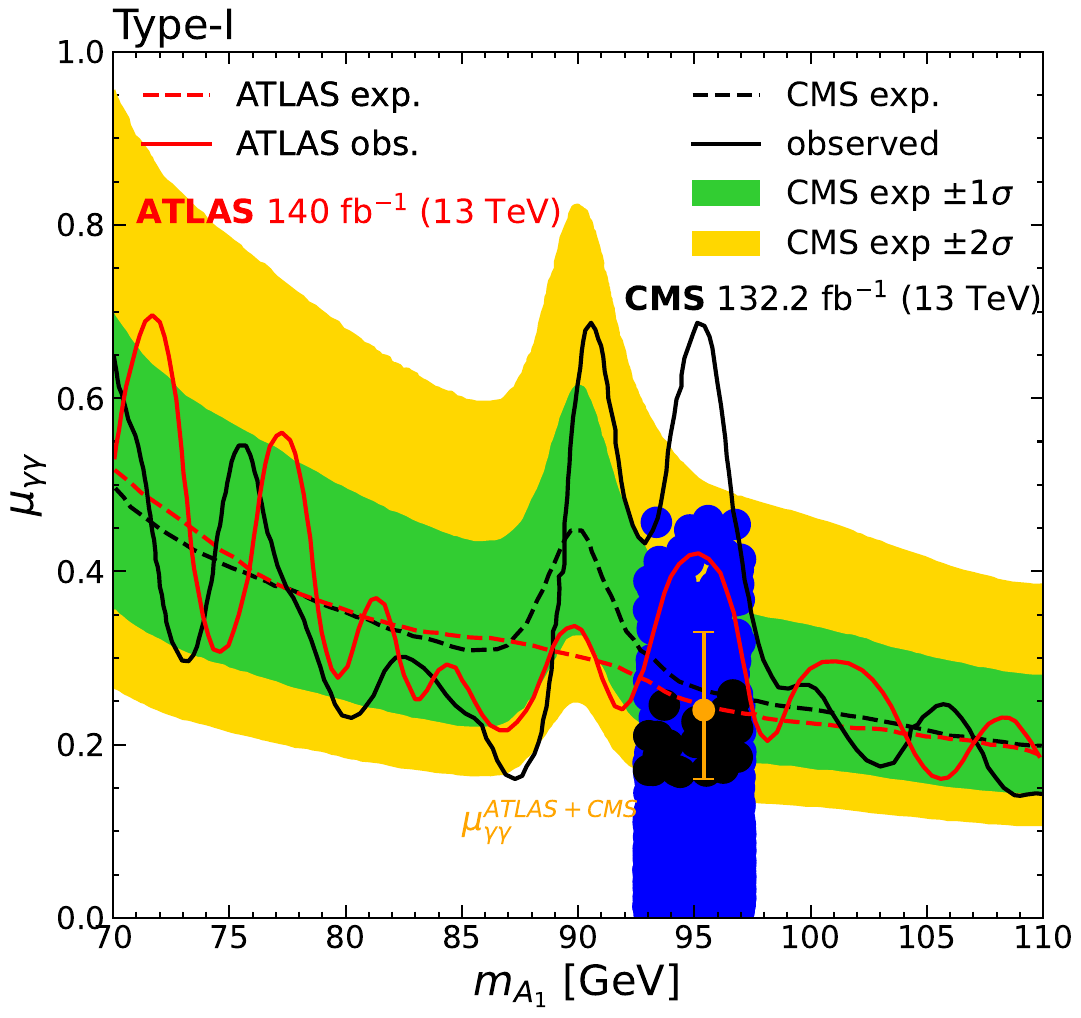}~
			\caption{\small Scattered points in the ($m_{h_1}$,$\mu_{\gamma\gamma}$) and ($m_{A_1}$,$\mu_{\gamma\gamma}$) planes. The dashed and solid lines represent the expected and observed cross-section limits obtained by CMS and ATLAS, respectively. CMS is depicted in black, while ATLAS is shown in red. The green and yellow bands represent the $1\sigma$ and $2\sigma$ uncertainty intervals, respectively. The yellow error bar indicates the values of $\mu_{\gamma\gamma}^{CMS+ATLAS}$, along with its uncertainty. The black points feature $\chi^2_{\gamma\gamma+\tau^+\tau^-+b\bar{b}} \leq 3.53$
				and describe the excesses at the level of $1\,\sigma$ or better, whereas the blue points represent $\chi^2_{\gamma\gamma+\tau^+\tau^-+b\bar{b}} > 3.53$.}
			\label{mugg}
		\end{figure*}
		\begin{figure*}[hbtp]
			\centering
			\includegraphics[width=0.40\textwidth]{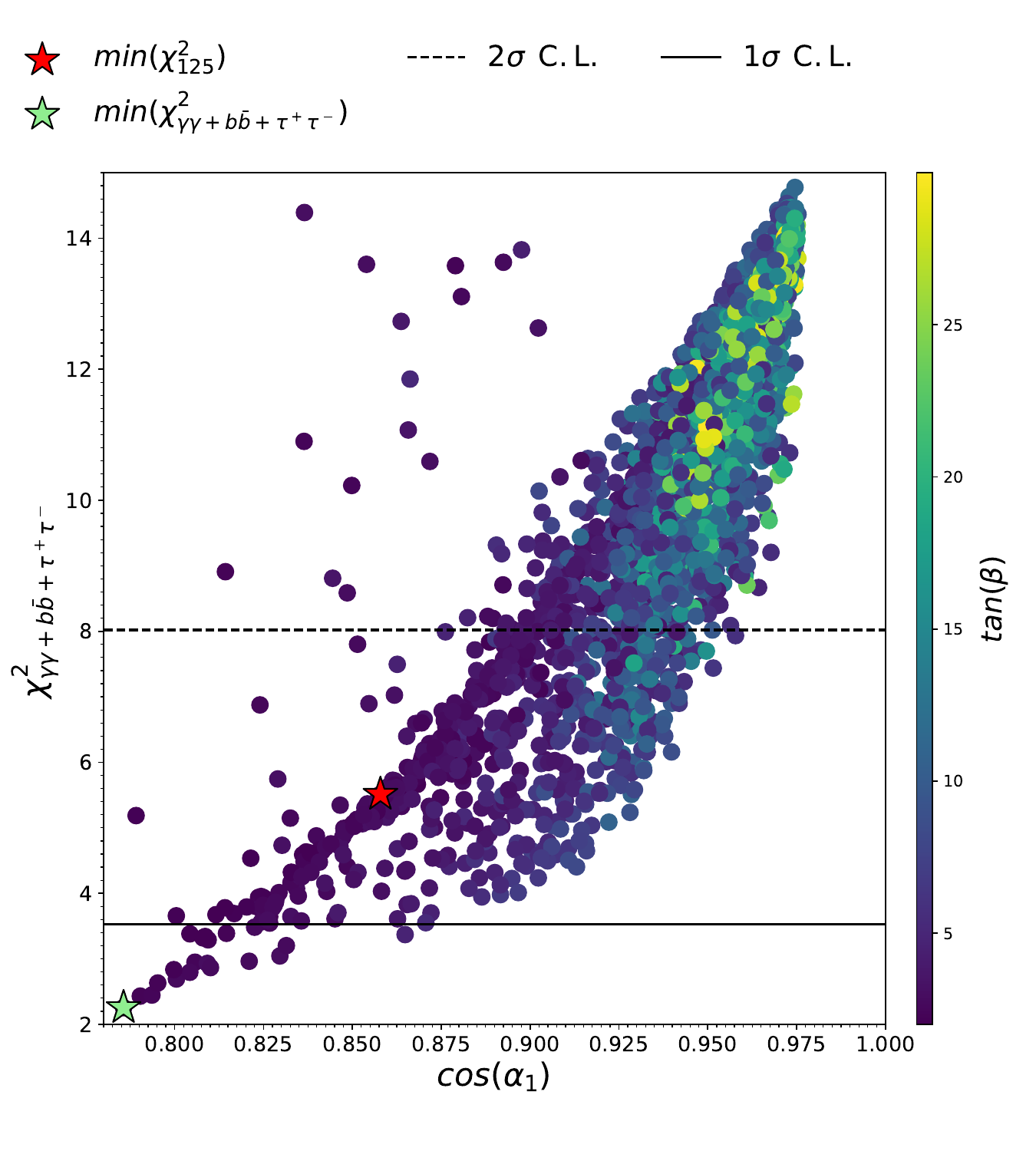}~
			\caption{\small
				Scattered points in the plane of ($cos\ \alpha_1$,$\ \chi^2_{\gamma\gamma+\tau^+\tau^-+b\bar{b}}$). The color bar indicates the value of $tan\beta$. The horizontal solid and dashed lines represents the $1\sigma$ and $2\sigma$ regions respectively.}
			\label{chi2}
		\end{figure*}
		\begin{table*}[t] 
			{\footnotesize	
				\setlength{\tabcolsep}{0.1cm}
				\renewcommand*{\arraystretch}{1.8}
				\begin{tabular*}{\textwidth}{ @{\extracolsep{\fill}} lcccccccccccccccccc}\noalign{\hrule height 0.8pt}
					\noalign{\vspace{1.25pt}}  
					\noalign{\hrule height 0.3pt}
					\textbf{Parameters} &
					
					$m_{h_1}$ & $m_{h_2}$ & $\lambda_1$ & $\lambda_3$ & $\lambda_4$ & $\lambda_6$ & $\lambda_7$ &  $\lambda_8$ & $\lambda_9$ & $\bar{\lambda_8}$ & $\bar{\lambda_9}$ & $tan\beta$ & $\alpha_1$ & $\alpha_2$ & $\alpha_3$ & $v_t$ & $\mu_{1}$&
					\\
					\noalign{\hrule height 0.9pt}
					&95.56 & 125.09 & 0.13 & 0.43 &   -0.13 &  2.99 &     4.06 &     1.35 &    2.63 &    2.64 &     -1.27 &     2.25 &     -0.66 &    -0.08 &     -0.02 &     1 &     -43.57& \\
					\noalign{\vspace{1.25pt}}  \noalign{\hrule height 0.9pt}
				\end{tabular*}
				{	\setlength{\tabcolsep}{0.1cm}
					\begin{tabular*}{\textwidth}{@{\extracolsep{\fill}}lccccccccccccccccc}
						\textbf{Signal strengths} &&&
						$\mu_{\gamma\gamma}(h_1)$ &&$\mu_{\gamma\gamma}(A_1)$&&
						$\mu_{\gamma\gamma}(h_1+A_1)$&&$\mu_{\tau^+\tau^-}(h_1)$&&$\mu_{\tau^+\tau^-}(A_1)$&&$\mu_{\tau^+\tau^-}(h_1+A_1)$&&$\mu_{b\bar{b}}(h_1)$\\
						\noalign{\hrule height 0.9pt}
						&&& 0.02&&   0.19    &&       0.21 && 0.45 &&0.19 && 0.65 && 0.062\\
						\noalign{\vspace{1.25pt}}  \noalign{\hrule height 0.9pt}
						
				\end{tabular*}}
				\caption{\small  Description of our point corresponding to $min(\chi^2_{\gamma\gamma+\tau^+\tau^-+b\bar{b}})$.}
				\label{description}
			}
		\end{table*}

		\section{Conclusions}
		\label{conclusion}
		In the present paper, driven by emerging observations of a scalar resonance with an invariant mass distribution about $95$ GeV, we have studied whether the 2HDM augmented with a complex scalar triplet, can account for the reported excesses. Our analysis and results satisfied  a full set of constraints on the model parameter space, with the $h_2$ Higgs predicted by the model spectrum mimics the observed $h_{125}$ at the LHC. First, we analyzed the two excesses in the $\gamma\gamma$ and $b\bar{b}$ decay channels. We showed that the 2HDMcT model with the Type-I Yukawa texture can simultaneously accommodate these two excesses. Then, we remarkably found that the three excesses in  $\gamma\gamma$, $b\bar{b}$ and $\tau^+\tau^-$ channels can well be explained simultaneously, reaching a $0.64 \sigma$ C.L. when the CP-odd Higgs boson $A_1$ is nearly mass degenerate and superposed to the light Higgs $h_1$.  In the near future, the Run 3 data from ATLAS and CMS around $95$ GeV will help shed further light and clarify whether the observed excesses are an early indication of a new particle discovery and new physics beyond the Standard Model.
		\section*{APPENDIX : The Higgs Decay to Two Photons}
		\label{appendice:A}
		In addition to the contributions from Standard Model particles, the loop-mediated decay of a scalar particle also receives extra contributions from the charged Higgs particles. In the context of the Two Higgs Doublet Model with a Triplet (2HDMcT), which includes an additional Triplet field, the particle spectrum predicts the existence of three pairs of charged particles. This is two more than what the conventional Two Higgs Doublet Model (2HDM) predicts. The decay width for $h_1\to\gamma\gamma$ in 2HDMcT is given by,
		\begin{align}
			\Gamma({h_1} \rightarrow\gamma\gamma) 
			&=  \frac{G_\mu\alpha^2 M_{{h_1}}^3}
			{128\sqrt{2}\pi^3} \bigg| \sum_{f=l,q} N_c Q_f^2 g_{{h_1} ff} A_{1/2}^{{h_1}}  (\tau_f) + g_{{h_1} VV} A_1^{{h_1}} (\tau_W)   
			+\sum_{i=1,2} \frac{M_W^2 \lambda_{{h_1} H_i^+ H_i^-} }{2c_W^2 M_{H_i^\pm}^2} A_0^{h_1}(\tau_{H_i^\pm})
			\nonumber \\ 
			&+\frac{2M_W^2 \lambda_{{h_1} H^{++} H^{--}} }{c_W^2 M_{H^{\pm\pm}}^2} A_0^{h_1}(\tau_{H^{\pm\pm}}) \bigg|^2 
			\label{hgg}
		\end{align}
		
		The reduced couplings $g_{h_1 ff}$ and $g_{h_1 VV}$ of the Higgs bosons to 
		fermions and $W$ bosons are given in Tab.~\ref{table1} and Tab.~\ref{table2}, while the trilinear 
		$\lambda_{h_1 H_i^+ H_i^-}$ and $\lambda_{h_1 H^\pm H^\pm}$ couplings to charged Higgs bosons are given by \cite{Djouadi:2005gj},
		\begin{align}
			A_{1/2}^{h_1}(\tau) & = 2 [\tau +(\tau -1)f(\tau)]\, \tau^{-2}  \nonumber \\   
			A_1^{h_1}(\tau) & = - [2\tau^2 +3\tau+3(2\tau -1)f(\tau)]\, \tau^{-2} \nonumber \\
			A_{0}^{h_1}(\tau) & = - [\tau -f(\tau)]\, \tau^{-2}
		\end{align}
		\noindent
		The scaling variables are defined as $\tau_i=M^2_{\Phi}/4M^2_i$, where $M_i$ represents the loop mass. In this context, $M_i$ encompasses all Standard Model charged particles as well as the Charged Higgs bosons predicted by the model, which provide additional contributions to the process.
		\begin{eqnarray}
			\bar{\lambda}_{h_1H^\pm_1H^\pm_1}&=&\frac{s_w}{2e m_w}\left(2 \mathcal{C}_{21}^2 \left(\lambda _6 \mathcal{E}_{13} v_{\Delta }+\lambda _1 v_1 \mathcal{E}_{11}+\lambda _3 v_2 \mathcal{E}_{12}\right)\right.\nonumber\\
			&+&2 \mathcal{C}_{22}^2 \left(\lambda _7 \mathcal{E}_{13} v_{\Delta }+\lambda _2 v_2 \mathcal{E}_{12}+\lambda _3 v_1 \mathcal{E}_{11}\right)\nonumber\\
			&+&\mathcal{C}_{23}^2 \left(2 \mathcal{E}_{13} (2\overline{\lambda _{8}}+\overline{\lambda _{9}}) v_{\Delta }+\left(2 \lambda _6+\lambda _8\right) v_1 \mathcal{E}_{11}+\left(2 \lambda _7+\lambda _9\right) v_2 \mathcal{E}_{12}\right)\nonumber\\
			&+&\mathcal{C}_{22} \mathcal{C}_{23} \left(\sqrt{2} \lambda _9 \mathcal{E}_{12} v_{\Delta }+\sqrt{2} \lambda _9 v_2 \mathcal{E}_{13}-4 \mu _2 \mathcal{E}_{12}-2 \mu _3 \mathcal{E}_{11}\right)\nonumber\\
			&+&\mathcal{C}_{21} \left(\right.\mathcal{C}_{23} \left(\sqrt{2} \lambda _8 \mathcal{E}_{11} v_{\Delta }+\sqrt{2} \lambda _8 v_1 \mathcal{E}_{13}-4 \mu _1 \mathcal{E}_{11}-2 \mu _3 \mathcal{E}_{12}\right)\nonumber\\
			&+&\left.2 \mathcal{C}_{22} \left(\lambda _4+\lambda _5\right) \left(v_2 \mathcal{E}_{11}+v_1 \mathcal{E}_{12}\right)\left.\right)\right)
			\label{3.3}
		\end{eqnarray}
		\begin{eqnarray}
			\bar{\lambda}_{h_1H^\pm_2H^\pm_2}&=&\frac{s_w}{2e m_w}\left(2 \mathcal{C}_{31}^2 \left(\lambda _6 \mathcal{E}_{13} v_{\Delta }+\lambda _1 v_1 \mathcal{E}_{11}+\lambda _3 v_2 \mathcal{E}_{12}\right)\right.\nonumber\\
			&+&2 \mathcal{C}_{32}^2 \left(\lambda _7 \mathcal{E}_{13} v_{\Delta }+\lambda _2 v_2 \mathcal{E}_{12}+\lambda _3 v_1 \mathcal{E}_{11}\right)\nonumber\\
			&+&\mathcal{C}_{33}^2 \left(2 \mathcal{E}_{13} (2\overline{\lambda _{8}}+\overline{\lambda _{9}}) v_{\Delta }+\left(2 \lambda _6+\lambda _8\right) v_1 \mathcal{E}_{11}+\left(2 \lambda _7+\lambda _9\right) v_2 \mathcal{E}_{12}\right)\nonumber\\
			&+&\mathcal{C}_{32} \mathcal{C}_{33} \left(\sqrt{2} \lambda _9 \mathcal{E}_{12} v_{\Delta }+\sqrt{2} \lambda _9 v_2 \mathcal{E}_{13}-4 \mu _2 \mathcal{E}_{12}-2 \mu _3 \mathcal{E}_{11}\right)\nonumber\\
			&+&\mathcal{C}_{31} \left(\right.\mathcal{C}_{33} \left(\sqrt{2} \lambda _8 \mathcal{E}_{11} v_{\Delta }+\sqrt{2} \lambda _8 v_1 \mathcal{E}_{13}-4 \mu _1 \mathcal{E}_{11}-2 \mu _3 \mathcal{E}_{12}\right)\nonumber\\
			&+&\left.2 \mathcal{C}_{32} \left(\lambda _4+\lambda _5\right) \left(v_2 \mathcal{E}_{11}+v_1 \mathcal{E}_{12}\right)\left.\right)\right)
			\label{3.4}
		\end{eqnarray}
		\begin{eqnarray}
			\bar{\lambda}_{h_1H^{\pm\pm}H^{\pm\pm}}&=&\frac{s_w}{e m_w}\left(2 \overline{\lambda _{8}} \mathcal{E}_{13} v_{\Delta }+\lambda _6 v_1 \mathcal{E}_{11}+\lambda _7 v_2 \mathcal{E}_{12}\right)
			\label{3.5}
		\end{eqnarray}	
		
		%
		
		\bigskip
		
		\noindent
		%
		%
		\section{Acknowledgments}
		
         The authors would like to thank CNRST/HPC-MARWAN for technical support.

\bibliography{references}{}
\bibliographystyle{JHEPcust}

\end{document}